\documentclass{llncs}
\usepackage[utf8]{inputenc}
\usepackage{siunitx}
\usepackage{amssymb}
\usepackage{amsmath}
\usepackage{graphicx}
\usepackage{url}
\usepackage{multirow}
\usepackage{subfig}
\usepackage{todos}
\usepackage{xspace}
\usepackage{graphicx}
\usepackage{booktabs}
\usepackage{fullpage}

\usepackage{hyperref}
\hypersetup{
  colorlinks=true,
  citecolor=blue,
  linkcolor=blue,
  urlcolor=blue,
  bookmarksnumbered,
  bookmarksopen,
  pdftitle={Transit Node Routing Reconsidered},
  pdfauthor={Julian Arz, Dennis Luxen, Peter Sanders},
  pdfsubject={Transit Node Routing Based on Contraction Hierarchies},
  pdfkeywords={graph, priority queue, Dijkstra, shortest path, contraction hierarchies, ch, transit node routing, distance oracle},
}

\newcommand{\etal}{et al.\xspace}
\newcommand{\ch}{Contraction Hierarchies\xspace}
\newcommand{\set}[1]{\left\{ #1\right\}}
\newcommand{\gilt}{:}
\newcommand{\setGilt}[2]{\left\{ #1\gilt #2\right\}}
\newcommand{\Vor}[1]{\mathrm{Vor}(#1)}

\title{Transit Node Routing Reconsidered\thanks{This work was partially supported by DFG Grant 933/2.}}
\author{Julian Arz, Dennis Luxen, Peter Sanders}
\institute{Karlsruhe Institute of Technology (KIT), 76128 Karlsruhe, Germany\\
 \email{\small julian.arz@student.kit.edu, \{luxen,sanders\}@kit.edu}
}
\date{}
\begin{document}

\maketitle

\begin{abstract}
Transit Node Routing (TNR) is a fast and exact distance oracle for road networks. We show several new results for TNR. First, we give a surprisingly simple implementation fully based on \ch that speeds up preprocessing by an order of magnitude approaching the time for just finding a \ch (which alone has two orders of magnitude larger query time). We also develop a very effective purely graph theoretical locality filter without any compromise in query times. Finally, we show that a specialization to the online many-to-one (or one-to-many) shortest path further speeds up query time by an order of magnitude. This variant even has better query time than
the fastest known previous methods which need much more space. 
\end{abstract}
\section{Introduction and Related Work}\label{sec:tnr-related}
Route planning in road networks has seen a lot of results from the algorithm engineering community in recent years.
With Dijkstra's seminal algorithm being the baseline, a number of techniques  preprocess the static input graph to achieve drastic speedups.
\textit{Contraction Hierarchies (CH)} \cite{gssd-chfsh-08,gssv-erlrn-12} is a speedup-technique that has a convenient trade-off between preprocessing effort and query efficiency.
Road network with millions of nodes and edges can be preprocessed in mere minutes while queries run in about a hundred microseconds.
Transit Node Routing (TNR) \cite{bfss-frrnt-07} is one of the fastest speed-up techniques for shortest path distance queries in road networks. 
By preprocessing the input road network even further, it yields \textit{almost} constant-time queries, in the sense that nearly all queries can be answered by a small number of table lookups.
It follows an intuition:
Long distance connections almost always enter an arterial network connecting a set of important nodes -- the \textit{transit nodes}.
The set of these entrances for a particular node is small on average. 
Once the \emph{transit nodes} are identified, a mapping from each node to its access nodes and pair-wise distances between all transit nodes is stored.
Preprocessing needs to compute a table of distances between the transit nodes, the distances to the access nodes, and information for a so-called \emph{locality filter}.
The filter indicates whether the shortest path might not cross any transit nodes, requiring an additional local path search. 

Many TNR variants have a common drawback that preprocessing time for TNR is significantly longer than for the underlying speed-up technique. 
Another weakness is that the locality filter requires geometric information on the position of the nodes \cite{bfm-trans-06,bfmss-itcsp-07,g-ch-08}. 
The presence of a geometric component in an otherwise purely graph theoretical method is regarded as awkward. 
There are several examples of geometric ingredients in routing techniques being superseded by more elegant and effective graph theoretical ones \cite{s-rprn-08,dssw-erpa-09} with the locality filter of TNR being the only \textit{survivor} that is still competitive.
Geisberger \cite{g-ch-08} uses CH to define transit node sets and for local searches, but still uses a geometric locality filter and relies on Highway Hierarchies \cite{ss-ehh-12} for preprocessing. 
In lecture slides \cite{Bast11}, Bast
describes a simple variant of CH-based preprocessing exploring a larger search space than ours which also computes a super-set of the access nodes because it omits post-search-stalling. No experiments are reported. The geometric locality filter is not touched. In Section~\ref{sec:tnr-our-variant} we remove all these qualifications and present a simple fully CH-based variant of TNR which yields surprisingly good preprocessing times and allows for a very effective fully graph-theoretical locality filter.

A related technique is \textit{Hub Labeling (HL)} by Abraham \etal \cite{adgw-ahbla-11} which stores sorted CH search spaces, intersecting them to obtain the distance. 
Using sophisticated tuning measures this can be made significantly faster than TNR since it incurs less cache faults. However HL need much more space than TNR. 

In Section~\ref{s:manyToOne} we further accelerate TNR queries for the special case that there are many queries with fixed target (or source). This method is even faster than HL without incurring its space overhead.

\section{Preliminaries}\label{sec:tnr-preliminaries}
We model the road network as a directed graph $G=(V,E)$, with $\vert E \vert=m$ edges and $\vert V\vert=n$ nodes.
Each node corresponds to a location, e.g. a junction, and edges represent the connections between them.
Each edge $e\in E$ has an associated cost $c(e)$, where $c:E\rightarrow \mathbb{R}^+$.
It is called the \textit{weight}.
A path $P=\langle s,v_1, v_2, \dotsc,t\rangle$ in $G$ is a sequence of nodes such that there exists an edge between each node and the next one in $P$.
The length of a path $c(P)$ is the sum its edge weights.
A path with minimum cost between $s,t\in V$ is called a \textit{shortest} path and denoted by $d(s,t)$ with cost $\mu(s,t)$. 
Note that a shortest path need not be unique.
A path $P=\langle v_0, v_1, \dotsc, v_p\rangle$ is called \textit{covered} by a node $v\in V$ if and only if $v \in P$. 
\subsection{Contraction Hierarchies}\label{sec:tnr-ch}
\ch heuristically order the nodes by some measure of importance and \emph{contract} them one by one in this order.
Contracting means that a node is removed (temporarily) and as few \emph{shortcut} edges as possible are inserted to preserve shortest path distances.

The CH search graph is the union of the set of original edges and the set of shortcuts with edges only leading to more important nodes.
This graph is a directed acyclic graph (DAG).
An important structural property of CHs is that for any two nodes $s$ and $t$, if there is an $s$--$t$-path at all, then there is also a shortest up-down path $s$--$m$--$t$ where $s$--$m$ uses only upward edges and $m$--$t$ uses only downward edges in the CH. The \emph{meeting} node $m$ is the highest node on this path in the CH.
The only crucial difference between a bidirectional Dijkstra and a CH query is the stopping criterion of the bidirected search.
It continues adding nodes into the priority queue until the tentative distances of added nodes exceed any upper bound that may exist for the shortest path.
The shortest path goes over a \emph{middle node} that is settled in both searches and for which CH guarantees correct labelling in both directions.

Although the search spaces explored in CH queries are rather small, there is a simple technique called \emph{stall-on-demand} \cite{s-rprn-08} that further prunes search spaces.
We use a simplified version of that technique, which leads to queries as fast as those reported in \cite{v-femno-10}.
For every node $v$ that is the end point of a relaxed edge $(u,v)$ it is checked if there exists a reverse edge $(w,v)$.
The edge is not relaxed if the tentative distance of $w$ plus the edge weight of $(w,v)$ is less than the distance of $u$ plus edge $(u,v)$.
If such a node $w$ exists, edge $(u,v)$ can't be part of a shortest path and thus $v$ is not added into the queue.
This is done by scanning the edges incident to $v$.

Computing a table of all pair-wise shortest path distances for a set of nodes can be done by running a quadratic number of queries.
While this is already significantly faster with CH than with a naive implementation of Dijkstra's algorithm, tables can be computed much more efficiently with the two-phase algorithm of Knopp \etal \cite{ksssw-cmmsp-07}.
Computing large distance tables is a matter of seconds since only $O(\vert S\vert + \vert T\vert)$ half searches have to be conducted.
The quadratic overhead to initialize and update table entries is close to none for $\vert S\cup T\vert = \mathcal{O}(\sqrt{n})$.
We refer the interested reader to \cite{ksssw-cmmsp-07,gssv-erlrn-12}.

\section{Transit Node Routing}
\label{sec:tnr-tnr}

TNR in itself is not a complete algorithm, but a framework.
A concrete instantiation has to find solutions to the following degrees of freedom:
It has to identify a set of transit nodes.
It has to find access node for all nodes.
And is has to deal with the fact that some queries between nearby nodes cannot answered via the transit nodes set.
In the remainder of this Section, we define and introduce the minimal ingredients for the generic TNR framework, conceive a concrete instantiation and then discuss an efficient implementation.

\begin{definition}
\label{tnr:dfn-tnr}
Formally, the generic TNR framework consists of
\begin{enumerate}
\item A set $\mathcal{T} \subseteq V $ of transit nodes.
\item A \textit{distance table} 
  $D_\mathcal{T}: \mathcal{T} \times \mathcal{T} \to \mathbb{R} _0^+$
  of shortest path distances between the transit nodes.
\item A forward (backward) \textit{access node mapping} $A^\uparrow : V   \rightarrow 2^\mathcal{T}$ ($A^\downarrow: V \rightarrow 2^\mathcal{T}$).  For any shortest $s$--$t$-path $P$ containing transit nodes, $A^\uparrow(s)$ $\left(A^\downarrow(t)\right)$ must contain the first (last) transit node on $P$.
\item A \textit{locality filter} 
  $\mathcal{L}:V\times V\rightarrow\{\text{true},   \text{false}\}$.   
  $\mathcal{L}(s,t)$ must be true when no shortest path   
  between $s$ and $t$ is covered by a transit node. 
  False positives are allowed, i.e., $\mathcal{L}(s,t)$ may sometimes 
  be true even when a shortest path
  contains a transit node.
\end{enumerate}
\end{definition}
Note that we use a simplified version of the generic TNR framework. 
A more detailed description is in Schultes' Ph.D. dissertation \cite{s-rprn-08}.
We outline a generalization to multiple layers of transit nodes in Section \ref{sec:tnr-query-time}.
During preprocessing $\mathcal{T}$, $D_\mathcal{T}$, $A^\uparrow$, $A^\downarrow$, and some information sufficient to evaluate $\mathcal{L}$ is precomputed. An $s$--$t$-query first checks the locality filter. If  $\mathcal{L}$ is true, then some fallback algorithm is used to handle the local query. Otherwise, 
\begin{equation}\label{eq:distance}
\mu(s,t)=\mu_{min}(s,t) :=  \mspace{-9mu}\min_{\substack{a_s\in A^\uparrow(s)\\a_t\in A^\downarrow(t)}} \{d_{A^\uparrow}(s,a_s)+D_\mathcal{T}(a_s,a_t)+d_{A^\downarrow}(a_t,t)\}.
\end{equation}

\section{CH based TNR}\label{sec:tnr-our-variant}
Our TNR variant  (CH-TNR) is based on CH and does not require any geometric information. We start by selecting a set of transit nodes. 
Local queries are implicitly defined and we find a locality filter to classify them. For simplicity, we assume that the graph is strongly connected. In Section~\ref{sec:tnr-query-time} we discuss what needs to be done to handle the general case.

\subsubsection{Selection of Transit Nodes.}
CHs order the  nodes in such a way that nodes occurring in many shortest paths are moved to the upper part of the hierarchy.
Hence, CH is a natural choice to identify a small node set which covers many shortest paths in the road network.
We chose a number of transit nodes $\vert\mathcal{T}\vert=k$ and select the highest $k$ nodes from the CH data structure.
This choice of $\mathcal{T}$ also allows us to exploit valuable structural properties of CHs. 
A distance table of pair-wise distances is built on this set with a CH-based implementation of the many-to-many algorithm of Knopp \etal \cite{ksssw-cmmsp-07}. 

\subsubsection{Finding Access Nodes.}
We only explain how to find forward access nodes from a node $s$. The computation of backward access nodes works analogously. We will show that the following simple and fast procedure works: Run a forward CH query from $s$. Do not relax edges leaving transit nodes. When the search runs out of nodes to settle, report the settled transit nodes. 

\begin{lemma}\label{lem:findAccess}
The transit nodes settled by the above procedure find a superset of the access nodes of $s$ together with  with their shortest path distance.
\begin{proof}

Consider a shortest $s$--$t$-path $P:=\langle s,\ldots, t\rangle$ that is covered by a node $u\in\mathcal{T}$.
Furthermore, assume that $u$ is the highest transit node on $P$. 
A fundamental property of CHs is that we can assume $P$ to consist of upward edges leading up to $u$ followed by downward edges to $t$. 
Moreover, the forward search of a CH query finds the shortest path to $u$. 
Thus, a CH query also finds a shortest path to the first transit node $v$ on $P$. 
It remains to show that the pruned forward search of CH-TNR preprocessing does not prune the search before settling $v$. 
This is the case since pruning only happens when settling transit nodes and we have defined $v$ to be the first transit node on $P$.


\end{proof}
\end{lemma}

The resulting superset of access nodes is then reduced using \emph{post-search-stalling} \cite{s-rprn-08}:
For all nodes $t_1,t_2 \in A^\uparrow(v)$, if $d_{A^\uparrow}(v,t_1)+D_\mathcal{T}(t_1,t_2)\leq d_{A^\uparrow}(v,t_2)$, discard access node $t_2$.

\begin{lemma}\label{lem:minimality}
Post-search-stalling yields a set of access nodes that is minimal in the sense that it only reports nodes that are the first transit node on some shortest path starting on $s$.
\begin{proof}
Consider a transit node $t$ that is found by our search which is not an access node for $s$, i.e., there is an access node $u$ on every shortest path from $s$ to $t$. 
By Lemma~\ref{lem:findAccess}, our pruned search found the shortest path to $u$ but did not relax edges out of $u$.
Hence, the only way $t$ can be reported is that it is reported with a distance larger than the shortest path length.
Hence, $t$ will be removed by post-search-stalling.
\end{proof}
\end{lemma}

\subsubsection{Search Space Based Locality Filter.}
Consider a shortest path query from $s$ to $t$.  Let $S_{\uparrow}(s)$ denote the sub-transit-node search space considered by a CH query from $s$, i.e., those nodes $v$ settled by the forward search from $s$ which are not transit nodes. Analogously, let $S_{\downarrow}(t)$ denote the sub-transit-node CH search space backwards from $t$. If these two node sets are disjoint, all shortest up-down-paths from $s$ to $t$ must meet in the transit node set and hence, we can safely set $\mathcal{L}(s,t)=\text{false}$. Conversely, if the intersection is non-empty, there might be a meeting node below the transit nodes corresponding to a shortest path not covered by a transit node. 

Thus a very simple locality filter can be implemented by storing the sub-transit-node search spaces which are computed for finding the access nodes anyway. 
\begin{lemma}\label{lem:searchspace}
The locality filter described above fulfils Definition~(\ref{tnr:dfn-tnr}).
\begin{proof}
We assume for $s, t \in V\backslash\mathcal{T}$, the distance $\mu(s, t) \neq \mu_{\min}(s, t)$ and thus $\mu(s, t) < \mu_{\min}(s, t)$. Then the meeting node $m$ of a CH-query is not a transit node, and it has to be in the forward search space for $s$, $\bar{S_\uparrow}(s)$ \emph{and} in the backward search space for $t$, $\bar{S_\downarrow}(t)$. 
Hence, $\bar{S_\uparrow}(s) \cap \bar{S_\downarrow}(t) \neq \emptyset$.
\end{proof}
\end{lemma}
%
Preliminary experiments indicate that the average size of these search spaces are much smaller than the full search spaces, e.g. 32 instead of 112 in the main test instance from Section~\ref{sec:tnr-experiments}.  For the locality filter, only node IDs need to be stored. Compared to hub labelling which has to store full search spaces and also distances to nodes this is already a big space saving. 

If we are careful to number the nodes in such a way that nearby nodes 
usually have nearby numbers, the node numbers appearing in a search space
will often come from a small range. We precompute and store these values
in order to facilitate the following \emph{interval check}: When
$[\min(\bar{S_\uparrow}(s))  ,\max(\bar{S_\uparrow}(s))]\cap
 [\min(\bar{S_\downarrow}(t)),\max(\bar{S_\downarrow}(t))]=\emptyset$, we 
immediately know that the search spaces are disjoint. As the sole locality filter this would allow too many false positives but it works sufficiently often to drastically reduce the average overhead for the locality filter.
Below we discuss a much more accurate lossy compression of the search spaces.

\subsubsection{Graph Voronoi Label Compression.}
Note that the locality filter remains correct when we add nodes to the search spaces. 
We do this by partitioning the graph into regions and define the extended search space as the union of all regions that contain a search space node. 
This helps compression since we can represent a region using a single id, e.g., the number of a node representing the region. 
This also speeds up the locality filter since instead of intersecting the search spaces explicitly, it now suffices to intersect the (hopefully smaller) sets of block ids.  
Hence, we want partitions that are large enough to lead to significant compression, yet small and compact enough to keep the false positive rate small. Our solution if a purely graph theoretical adaptation of a geometric concept.
Our blocks are \emph{graph Voronoi regions} of the transit nodes. Formally, $$\Vor{v}:=\setGilt{u\in V}{\forall   w\in\mathcal{T}\setminus\set{v}\gilt\mu(u,v)\leq \mu(u,w)}$$ for $v\in\mathcal{T}$ with ties broken arbitrarily. The intuition behind this is that a positive result of the locality filter means that the search spaces of start and destination come at least close to each other.
Computing the Voronoi regions is easy, using a single Dijkstra run with multiple sources on the reversed input graph, as shown by Mehlhorn \cite{m-as-88}. 
We call this filter the \emph{graph Voronoi filter}.

\section{Experimental Evaluation}
\label{sec:tnr-experiments}

We implement our algorithms and data structures in C++ and test the performance on a real-world data set.
The source code is compiled with GCC 4.6.1 setting optimization flags \texttt{-O3} and \texttt{mtune=native}.
Our test machine is an Intel Core i7-920, clocked at 2.67 GHz with four cores and 12 GiB of RAM.
It runs Linux kernel version 2.6.34

Our CH variant implements the shared-memory parallel preprocessing algorithm of Vetter \cite{v-ptdch-09} with a hop limit of $5$ and $1\,000$ settled nodes for witness searches and $7$ hops or $2\,000$ settled nodes during the actual contraction of nodes.
The priority function is $$2*\text{edgeQuotient} + 4*\text{originalEdgeQuotient} + \text{nodeDepth}.$$

We experiment on the road network of Western Europe provided for the 9th DIMACS challenge on shortest paths \cite{dgj-spndi-09} by PTV AG.
Results for further instances can be found in Section~\ref{sec:tnr-otherInstances}.
The graph consists of about 18\,015\,449 nodes and 22\,413\,128 edges with travel time metric weights. The resulting hierarchy has 39\,256\,327 edges.
The following experiments are conducted with a transit node set of size 10\,000, if not mentioned otherwise, because key results from previous work were based on the same number of transit nodes, e.g. \cite{bfmss-itcsp-07}.

We use two arrays to store all access nodes.
Array $A$ for the access nodes and distances, and an index array $I_A$. 
For each node $v$, $A$ contains two sets of entries, one for $A^\uparrow(v)$ and one for $A^\downarrow(v)$. For each access node $a \in A^\uparrow(v)$ (or $\in A^\downarrow(v)$), two values are stored, the ID of $a$ and the distance $d_{A^\uparrow}(v, a)$ (or $d_{A^\downarrow}(a, v)$).

The access nodes are ordered by ID, which leads to a better query cache efficiency.
The index array $I_A$ stores for each node the starting indices of its two access node sets in $A$.
At the end of the index array, a dummy value points to the index after the last value in $A$.
The search space based locality filters (with or without compression) are stored the same way, using arrays $S$ and $I_S$.

The following design choices are used throughout the experiments.
Forward and backward search spaces are merged into one set for the locality filter. 
Forward and backward access node sets are also merged into one set.
But note that these two sets are distinct in our implementation.

As the ID of a node does not contain any particular information, node IDs can be changed to gain algorithmic advantages. 
This \textit{renumbering} is done by applying a bijective permutation on the IDs, in order to ensure that each ID stays unique.
We alter the labels of the nodes in $V$ so that $\mathcal{T} = \{0, \dotsc, k-1\}$.
By proceeding this way, we can easily determine if a node $v$ is a transit node or not (during further preprocessing and during the query): $v \in \mathcal{T}$ if and only if $v < k$.

\paragraph{Node Renaming.}
We examine renumbering strategies separately for the transit nodes set  $\mathcal{T}$ and for the remaining part of the CH search graph $V\backslash\mathcal{T}$.
We follow two aims for renumbering here.
One is to make table lookups faster for non-local queries, while the other aim is to make local queries as fast as possible.
Therefore, we treat both parts of our data structure with different strategies.
Second, the numbering makes a difference on the performance of table lookups. 
If the access nodes of a node are from a more compact interval, the cache efficiency of the table lookups is increased. 
Consider a number of access nodes $\vert A^\uparrow(s)\vert=a$ and $\vert A^\uparrow(t)\vert=b$ for source and target nodes respectively.
Wlog, we assume $a<b$.
The obvious worst case is $a \cdot b$ cache misses, while the best case is $a$ misses only.
This happens when all $b$ entries are in one cache line. 

Table~\ref{tab:RenamingTransit} shows the impact of renumbering the transit node set. 
The strategy is \emph{input level}-based.
We iterate through each level of the hierarchy top-down and order the nodes in each level with respect to their order in the  input data.
In other words, the partial order in each level respects the order of the input data.

\begin{table}[b]
\caption{Impact of renumbering on $\mathcal{T}$. The remainder of the graph is in increasing DFS ordering.}
\label{tab:RenamingTransit}
\centering
\begin{tabular}{lrrrr}
\toprule
Transit & LF & Interval & TL & Total \\
Nodes & [\si{n\second}] & Test [\%] & [\si{\micro\second}] & [\si{\micro\second}] \\
\midrule
7000 	& 122 & 84.9 & 1.03 & 1.50 \\
14000 & 111 & 90.1 & 1.05 & 1.32 \\
21000 & 105 & 92.8 & 1.04 & 1.25 \\
28000 & 104 & 94.5 & 1.04 & 1.22 \\
\bottomrule 
\end{tabular}
\end{table}

\begin{table}[tb]
\caption{Different renumbering strategies for $V \setminus \mathcal{T} $. $\mathcal{T} $ is in input order.}
\label{tab:RenamingNonTransit}
\centering
\begin{tabular}{lrrr}
\toprule
 &	 & \multicolumn{2}{c}{Query} \\
 		\cmidrule(lr){2-2} \cmidrule(lr){3-4}
Preprocessing & dur & LS & Total\\
Strategy & [s] & [$\mu$s] & [$\mu$s]\\
\midrule
(greedy) DFS Increasing	& 16.9 	& 27.4 & 1.38 \\  
(greedy) DFS Decreasing	& 16.9 	& 32.2 & 1.41 \\ 
Input Level Ordering	& 8.9 & 38.4 & 1.45 \\ 
\bottomrule 
\end{tabular}
\end{table}

The lower, non-transit node portion of the CH search graph is also renumbered.
It is used for local queries only and thus it has no effect on non-local queries.
It also doesn't influence the preprocessing in our experiments and we attribute that to the fact that search spaces are similar when nodes are close to each other and the fact that input ordering already exhibits a \emph{good} locality.
Table~\ref{tab:RenamingNonTransit} gives results on different renumbering strategies that we detail in the following.
The \emph{(greedy) DFS} orderings renumber the graph according to a (modified) depth-first graph traversal (DFS), while the input DFS ordering preserves the partial ordering of the levels as described above for the transit node set.
For every node an upward DFS is conducted that relaxes edges in the CH search graph that lead to more important nodes.
More specifically nodes in $V \setminus \mathcal{T}$, which are not yet renumbered, are explored.
The actual renumbering happens during  the backtracking step, i.e. we renumber a node if and only if all of its successors are already renumbered. 
The actual IDs can be assigned in increasing ($0,2,\ldots,n-k-1$) or decreasing order ($n-k-1,\ldots,1,0$).
Column \emph{dur} gives the duration of the renumbering, while \emph{LS} gives the average running time of a local search.
Column \emph{total} gives the average running time over all queries.
We see from the results that the DFS strategy with increasing IDs works best with respect to efficiency of local queries.

During the search for the access nodes, stall-on-demand is used to decrease the search space sizes. 
We tested different variants, varying in the number of hops the stalling does look ahead to find a witness for a wrong distance. 
A higher number of hops on the one hand increases preprocessing time, but on the other hand decreases the search spaces, speeding up the locality filter construction.
Table~\ref{tab:StallOnDemand} shows that while an increase of the hop depths from 1 to 2 manages to decrease space overhead and query times, a further increase from 2 to 3 is inadvisable: 
The preprocessing takes \SI{43}{\minute} longer, and does only have slightly better results.
Local searches take less time with higher hop depths. 
This is an interesting observation, because the stalling during preprocessing should not affect local searches. 
An explanation is that the (due to a more exact locality filter) omitted local searches have a higher distance.

\begin{table}[b]
\caption{Different hop depths for the stall-on-demand during preprocessing.}
\label{tab:StallOnDemand}
\centering
\begin{tabular}{lrrrrrrr}
\toprule
 	& \multicolumn{4}{c}{Preprocessing} & \multicolumn{3}{c}{Query} \\
 		\cmidrule(lr){2-5} \cmidrule(lr){6-8}
	& Expl.	&  & Voronoi & Total & LS & LS & Total\\
Hops & [s] & $|\bar{S}|$ & $|\bar{S}|$ & [byte / node] & [\%] & [$\mu$s] & [$\mu$s]\\
\midrule 
0	& 301			& 93.0 			& 29.3 			& 296 		& 2.36 & 30.4 & 2.15 \\
1 	& \textbf{149} 	& 31.8 			& 8.0 			& 211 		& 0.58 & 27.4 & 1.38 \\ 
2	& 446 			& 28.1 			& 6.3 			& 204 		& 0.41 & 26.0 & 1.35 \\ 
3 	& 3237 			& \textbf{27.9} & \textbf{6.1} & \textbf{203} & \textbf{0.40} & \textbf{25.7} & \textbf{1.32} \\ 
\bottomrule 
\end{tabular}
\end{table}

$\mathcal{T}$ is renumbered with the so-called \textit{input-level strategy}, while $V\backslash\mathcal{T}$ is ordered by the (greedy) DFS Increasing strategy.
The interval check accelerates the average running time of the locality filter. 
Prior to running the merging step, we check in constant time if the two intervals overlap or not with the interval check. 

\paragraph{Scalability.}
We test the scalability of parallel preprocessing for a varying number of cores in
Table~\ref{tab:MultiThreadPreprocessing}. 
The raw results of parallelizable parts (preprocessing, distance table generation and exploration) have a quite high variance of about 10\%. 
Hence, we measured the preprocessing five times and averaged over all runs. 
The values reported in column \emph{Total} are the sum of the respective averages.
Column \emph{Cores} gives the numbers of cores used.
Columns \emph{CH}, \emph{Dist. Table}, \emph{Exploration} measure time, speedup and efficiency of the respective parts.
The bottom line reports on four CPUs with activated hyper-threading (HT).

\begin{table}[t]
\caption{Scalability Experiment with 10\,000 transit nodes.}
\label{tab:MultiThreadPreprocessing}
\centering
\begin{tabular}{lcrrrcccrrrcrrrcrrr}
\toprule
Cores & & \multicolumn{3}{c}{CH} & & \multicolumn{3}{c}{Dist. Table} & & \multicolumn{3}{c}{Exploration} &  & \multicolumn{3}{c}{Total} \\[0.5em]\cline{1-1}\cline{3-5}\cline{7-9}\cline{11-13}\cline{15-17}
& & [s] & Spdp & Eff. & & [s] &  Spdp & Eff. & & [s] & Spdp & Eff. & & [s] & Spdp. & Eff. \\
\midrule
1 & & 513 & 1 & 1 & & 9.0 & 1 & 1 &  & 500 & 1 & 1 & & 1046 & 1 & 1\\
2 & & 281 & 1.83 & 0.91 & & 5.1 & 1.74 & 0.88 & & 287 & 1.74 & 0.87 &  & 596 & 1.75 & 0.88 \\
3 & & 203 & 2.53 & 0.84 & & 3.9 & 2.23 & 0.76 & & 202 & 2.48 & 0.83 & & 432 & 2.42 & 0.81 \\
4 & & 160 & 3.20 & 0.80 & & 2.9 & 3.16 & 0.79 & & 145 & 3.43 & 0.86 & & 334 & 3.13 & 0.78 \\
\midrule
4 (HT) & & 137 & 3.75 & 0.47 & & 2.2 & 4.01 & 0.50 & & 101 & 4.93 & 0.62 & & 265 & 3.95 & 0.49 \\
\bottomrule 
\end{tabular}
\end{table}
We see that the total preprocessing time is only about a factor of two larger than plain CH preprocessing. Most additional work is due to search space exploration from each node.
We see that the different parts of the algorithm scale well with an increasing number of cores.
The total efficiency is slightly lower than the efficiency of the individual parts, as it includes about 23.6 seconds of non-parallelized work due to the Voronoi computation.
It does not reflect the performance of real cores, but HT comes virtually for free with modern commodity processors.
The rate of local queries is only 0.58~\%.
On average a non-local query takes 1.22~$\mu$s, while a local query takes 28.6~$\mu$s on average.
This results in an overall average query time of 1.38~$\mu$s and the space overhead amounts to 147 Bytes per node.


We compare to previous approaches to distance oracles for our test instance.
Some of these implementations were tested on an older AMD machine \cite{s-rprn-08} that was available for running the queries.
Table~\ref{tab:tnr-compare-tnr-variant} shows \textit{Reported} as given in the respective publications denoted by \emph{From}, while column \textit{Compared} gives running times either done on or normalized to the aforementioned AMD machine.
Therefore, similar to the methodology in \cite{bdsssw-chgds-10}, a scaling factor of 1.915 is determined by measuring preprocessing and query times on both machines using a smaller graph (of Germany).
Scaled numbers are indicated by a star symbol.
Values for CH were measured with our implementation.

The simplest TNR implementation is GRID-TNR that splits the input graph into grid cells and computes a distance table between the cells border nodes.
Note that the numbers for GRID-TNR were computed on a graph of the USA, but the characteristics should be similar to our test instance.
Preprocessing is prohibitively expensive while the query is about 20 times slower than ours.
The low space consumption is due to the fact that it is trivial to construct a locality filter for grid cells.
For HH-TNR \cite{s-rprn-08} and TNR+AF \cite{bdsssw-chgds-10}, preprocessing is single-threaded. 
The corresponding scaling factor for preprocessing is 3.551 and the fastest HH based TNR variant is still slower by about a factor of two for preprocessing and queries.
Note that the HH-based methods all implement a highly tuned TNR variant with multiple levels that is much more complex than our method.
While TNR+AF has faster queries by about 25\%, the (scaled) preprocessing is about an order of magnitude slower and the space overhead is twice as much.
Also, TNR+AF requires a sophisticated implementation with a partitioning step and the computation of arc flags.

\begin{table}[hb]
\caption{Comparison Between Various Distance Oracles.}
\label{tab:tnr-compare-tnr-variant}
\centering
\begin{tabular}{lcrrrrr}
\toprule 
		& 		& \multicolumn{2}{c}{Preprocessing} 	& \multicolumn{2}{c}{Query} \\
\cmidrule(lr){3-4} \cmidrule(lr){5-6}
		& 		& Reported  		& Space		& {Reported} & {Compared} \\
Method 	& From	& [min] & [byte / node] 	& {[$\mu$s]} & {[$\mu$s]}\\ 
\midrule
CH 				   & - 						& 2.7 	& 24 & 103 & 246 \\[5pt]
Grid-TNR 		   & \cite{bfmss-itcsp-07} 	& 1200 	&   21	& 63	 & 63 \\
HH-TNR-eco 		   & \cite{s-rprn-08}		& 25  	&  120  & 11  	 & 11 \\
HH-TNR-gen 		   & \cite{s-rprn-08} 		& 75  	&  247  &  4.30	 & 4.30 \\
TNR+AF 			   & \cite{bdsssw-chgds-10}	& 229 	&  321  &  1.90	 & 1.90 \\
HL local		   & \cite{adgw-ahbla-11}   & 159 	& 1221  &  0.572 & 1.10 & $\star$\\
HL global		   & \cite{adgw-ahbla-11}   & 165 	& 1269  &  0.276 & 0.53 & $\star$\\
HL-0 local 		   & \cite{adgw-hhlsp-12} 	&   3 	& 1341  &  0.7   & 1.34 & $\star$\\
HL-$\infty$ global & \cite{adgw-hhlsp-12} 	& 372 	& 1055  &  0.254 & 0.49 & $\star$\\
\midrule
CH-TNR             &	 -					& 5   	& 147 	&  1.38  & 3.27 \\
\bottomrule 
\end{tabular}
\end{table}

While the hub labeling based methods achieve superior query times, the reader should note the high space overhead incurred by these methods.
Even the most space efficient HL needs more than seven times higher more space.
HL-0 local reports faster preprocessing than our method with nine times higher space overhead.
It should be noted that these experiments were done on three times as many cores with 20\% faster clock speed of 3.2 GHz and 50\% larger L3 cache of 16 MiB.
Single core preprocessing for HL-O local takes 17.9 minutes while our approach is slightly faster with 17.4 minutes on a slower machine.
We acknowledge, though, that even HL with the fastest preprocessing has faster queries than ours by about a factor of 2--3. 
We attribute that to the higher number of cache misses of our method.

The quality of our locality filter is compared to other TNR implementations in Table~\ref{tab:LocalityFilter}.
These variants differ in the number of transit nodes and in the graph used to determine them.
Nevertheless, the graphs are road networks that exhibit similar characteristics.
The number of transit nodes for CH-TNR is chosen to resemble data from literature.
\begin{table}[t]
\caption{Comparison of Locality Filter Quality.}
\label{tab:LocalityFilter}
\centering
\begin{tabular}{lccrSSSS}
\toprule 
Method 		& {From} & &  $|\mathcal{T}|$  & Local  & {False} &\\ 
	 		&             &     & 		  &  [\%] & [\%] \\ 
\midrule
Grid-TNR	& \cite{bfmss-itcsp-07}	& &  7\,426 & 2.6  & \multicolumn{1}{c}{-} \\
Grid-TNR	& \cite{bfmss-itcsp-07}	& & 24\,899 & 0.8  & \multicolumn{1}{c}{-} \\
LB-TNR		& \cite{ef-tnlbr-12}   	& & 27\,843 & \multicolumn{1}{c}{-} & \multicolumn{1}{c}{-} \\
HH-TNR-eco 	& \cite{s-rprn-08}  	& &  8\,964 & 0.54 & 81.2 \\
HH-TNR-gen 	& \cite{s-rprn-08}		& & 11\,293 & 0.26 & 80.7 \\\hline
CH-TNR 		&\multicolumn{1}{c}{-} 	& & 10\,000 & 0.58 & 73.6 \\
CH-TNR		& \multicolumn{1}{c}{-} & & 24\,000 & 0.17 & 72.1 \\
CH-TNR		& \multicolumn{1}{c}{-} & & 28\,000 & 0.14 & 72.1 \\
\bottomrule 
\end{tabular}
\end{table}
We see that the fraction of local queries of our variant is lower than or on par with the numbers from literature.
Also, the rate of false positives is much lower than previous work.
Most noteworthy, the recent method of LB-TNR applies sophisticated optimization techniques, but does not produce a transit node set with superior locality as the rate of local queries is virtually the same.

\subsection{Impact of $\vert\mathcal{T}\vert$ on Query Efficiency}
\begin{figure}[tb]
    \centering
    \includegraphics[width=0.75\textwidth]{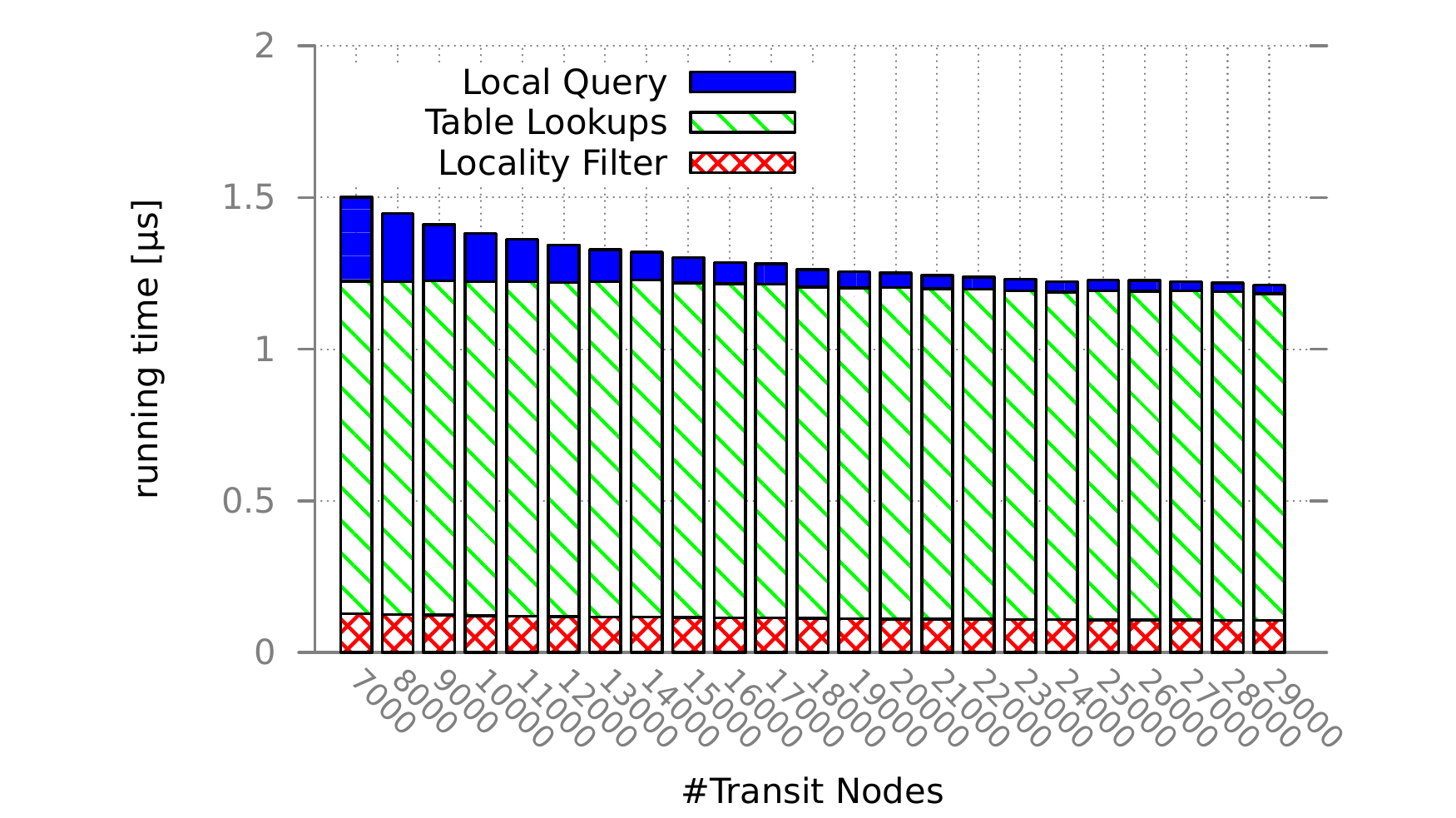}
    \caption{Average query times for different numbers of transit nodes.  Reported time for the local queries is averaged over the total number of queries.}
    \label{fig:plotRunningTime}
\end{figure}
Figure~\ref{fig:plotRunningTime} gives a row-stacked plot that details the contributions of each part of the query to the average total query time depending on the transit node set size.
We see that most of the query time is spent in table lookups and that this portion stays relatively stable over the entire parameter space.
We attribute that to two reasons.
First, the number of access nodes is relatively stable.
It drops from roughly 8.5 to just below 7.
Hence, the number of table lookups is also relatively stable.
Second, the table lookups are mostly dominated by cache misses and it appears that our renumbering effort is successful in that it already minimizes the number of cache misses across the board.

\begin{table}[hbt]
\caption{Average query time, preprocessing time (on four cores), and space overhead using a varying number of transit nodes.}
\label{tab:overallInfo}
\centering
\begin{tabular}{lccccccccc}
\toprule 
	& & Preproc. & & \multicolumn{2}{c}{Non-Local} & & \multicolumn{2}{c}{Local} & Amortized \\\cline{5-6}\cline{8-9}
$\vert\mathcal{T}\vert$ & & [\si{\minute}] & & [\%] & [\si{\micro\second}] & & [\%] & [\si{\micro\second}] & [\si{\micro\second}] \\
\midrule
7000  & & 6.1 & & 99.12 & 1.225 & & 0.88 & 32.661 & 1.501 \\
14000 & & 5.5 & & 99.62 & 1.229 & & 0.38 & 25.187 & 1.320 \\
21000 & & 5.2 & & 99.79 & 1.198 & & 0.21 & 21.977 & 1.243 \\
28000 & & 5.1 & & 99.86 & 1.189 & & 0.14 & 21.728 & 1.218 \\
\bottomrule 
\end{tabular}
\end{table}

In the following, preprocessing running times are reported for 4 threads.
As reported before, the size of the transit node set is a tuning parameter.
We look into the impact of varying the size of this set in the following experiments.
Especially, we explore the effect of transit node set size on the fraction of local queries, space overhead and query time.
Table~\ref{tab:overallInfo} reports on these experiments. 
Column $\vert\mathcal{T}\vert$ gives the size of the transit node set, while column \emph{Prepoc.} reports on the duration of the preprocessing.
Columns \emph{(Non-)Local} give the fraction of (non-)local queries and the respective query times.
Column \emph{Amortized} reports amortized query times, while \emph{Overhead} reports on the overhead per node.

The results of the experiment support the results from the previous section that the number of local queries decreases with an increasing transit node set.
This is expected behavior.
Likewise, the amortized query time decreases as the number of local queries drops.
This is also reflected in the absolute numbers of Table \ref{tab:LocalQueries} in which 1\,000\,000 random queries are performed.
\begin{table}[b]
\caption{The impact of $\vert\mathcal{T}\vert$ on the number of performed local queries.}
\label{tab:LocalQueries}
\centering
\begin{tabular}{lrrrr}
\toprule
				& \multicolumn{3}{c}{Local Searches} \\
\cmidrule{2-4}
	&				& Time					& Amortized \\					
$\vert\mathcal{T}\vert$	& \#Performed 	& [\si{\micro\second}] 	& [\si{\micro\second}] \\ 	
\midrule
7\,000 & 8\,798 & 31.4 & 0.277 \\
14\,000 & 3\,820 & 24.0 & 0.092 \\
21\,000 & 2\,128 & 20.8 & 0.044 \\
28\,000 & 1\,441 & 20.5 & 0.029 \\
\bottomrule 
\end{tabular}
\end{table}

\begin{table}[h]
\caption{Fraction of local queries according to the locality filter by Dijkstra Rank. Bold values show approximate $50\%$ threshold.}
\label{tab:filterByRank}
\centering
\begin{tabular}{lrrrrrrrrrrrrrrrrrrrr}
\toprule 
$\vert\mathcal{T}\vert$ & $\leq 2^9$ & $2^{10}$ & $2^{11}$ & $2^{12}$ & $2^{13}$ & $2^{14}$ & $2^{15}$ & $2^{16}$ & $2^{17}$ & $2^{18}$ & $2^{19}$ & $2^{20}$ & $\geq 2^{21}$ \\ 
\midrule
7\,000 & 100 & 100 & 99 & 98 & 96 & 88 & 74 & \textbf{56} & 34 & 15 & 5 & 1 & 0 \\
14\,000 & 100 & 99 & 98 & 95 & 86 & 68 & \textbf{47} & 28 & 12 & 4 & 1 & 0 & 0 \\
21\,000 & 100 & 99 & 96 & 89 & 73 & \textbf{51} & 29 & 14 & 5 & 1 & 0 & 0 & 0 \\
28\,000 & 100 & 97 & 93 & 81 & \textbf{61} & 38 & 19 & 8 & 2 & 1 & 0 & 0 & 0 \\
\bottomrule 
\end{tabular}
\end{table}

The decrease in local queries is not uniform across all Dijkstra ranks.
There appears to be a threshold after which the fraction of detected local queries falls sharply.
Table~\ref{tab:filterByRank} reports on the fraction of local queries that are performed depending on the Dijkstra rank.

We observe that increasing the transit node set size effectively lowers the rank when the locality filter detects roughly half of the queries as local queries.
These values are given in bold.
We see that its rank decreases by several orders of magnitude over the parameter space.

\begin{figure}[bht]
\centering
    \includegraphics{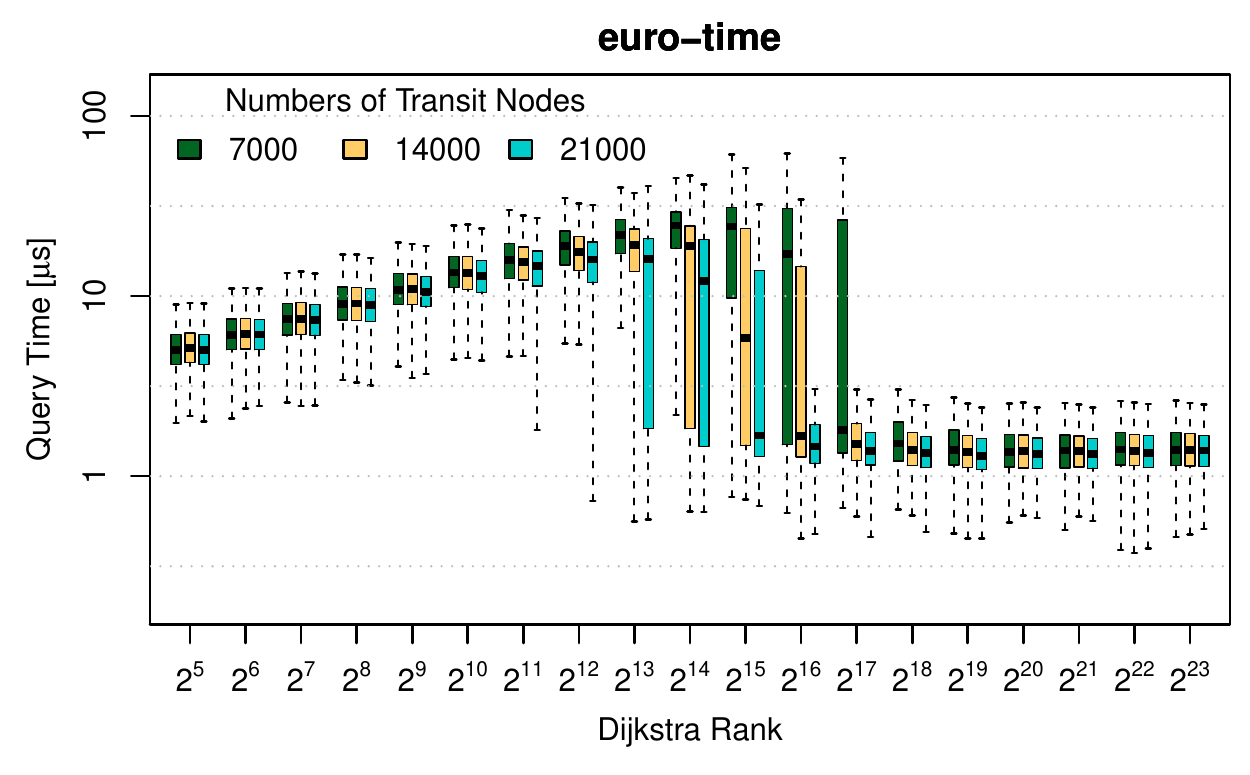}
    \caption{Rank plot for using $\vert\mathcal{T}\vert=\{7\,000, 14\,000,21\,000\}$ transit nodes.}
    \label{fig:rankPlots_time}
\end{figure}

A closer look at the query performance according to Dijkstra rank is given in Figures \ref{fig:rankPlots_time}.
The rank of a node $v$ with respect to node $s$ is $i$ if $v$ is the $i$-th node settled by a unidirectional Dijkstra query.
For sake of clear arrangement the Dijkstra rank $k:=\lfloor\log_2(i)\rfloor$ is the floored logarithm to base $2$ of $i$.
In other words, it gives a notion of distance independent of the graphs underlying geometry.

We see that the query time is dominated by the rather expensive fall-back algorithm for short-range queries in all the experiments.
Also, we see that the query time falls shortly for medium to long range queries once the shortest paths get covered by the transit node set.
Beyond this threshold the time approaches the bare minimum needed for running the locality filter and the table lookups, which is constant in practice.

\subsection{Impact of $\vert\mathcal{T}\vert$ on Space Overhead}
\begin{figure}[bth]
\centering
\subfloat[][Average number of nodes in the search spaces and corresponding number of Voronoi representatives.]{
\includegraphics[width=0.5\textwidth]{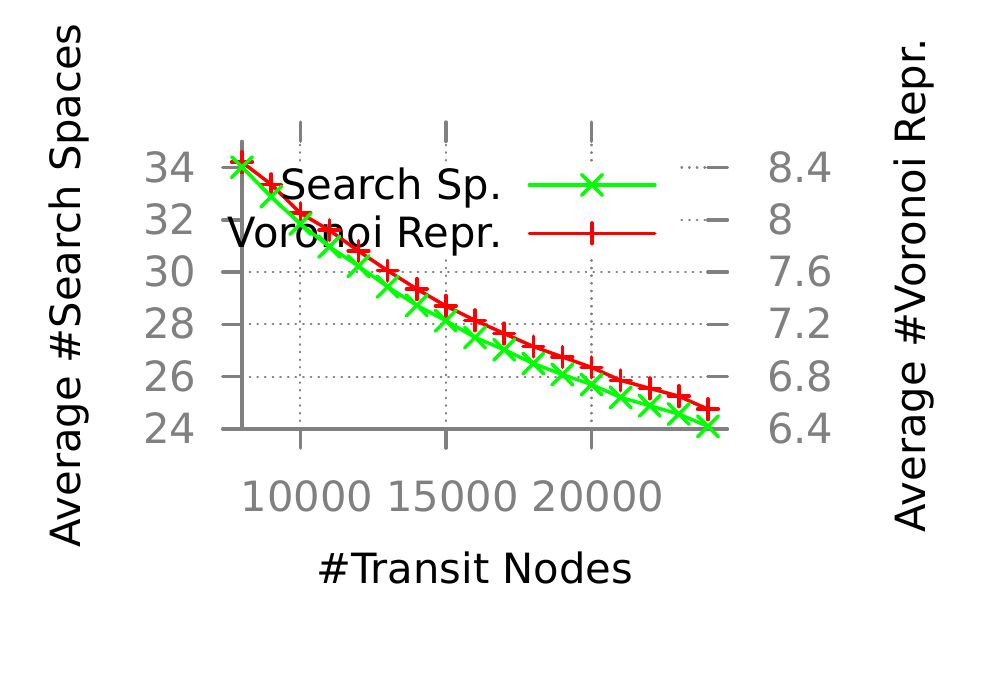}
\label{fig:plotSS}}
\subfloat[][Average number of access nodes.]{
\includegraphics[width=0.5\textwidth]{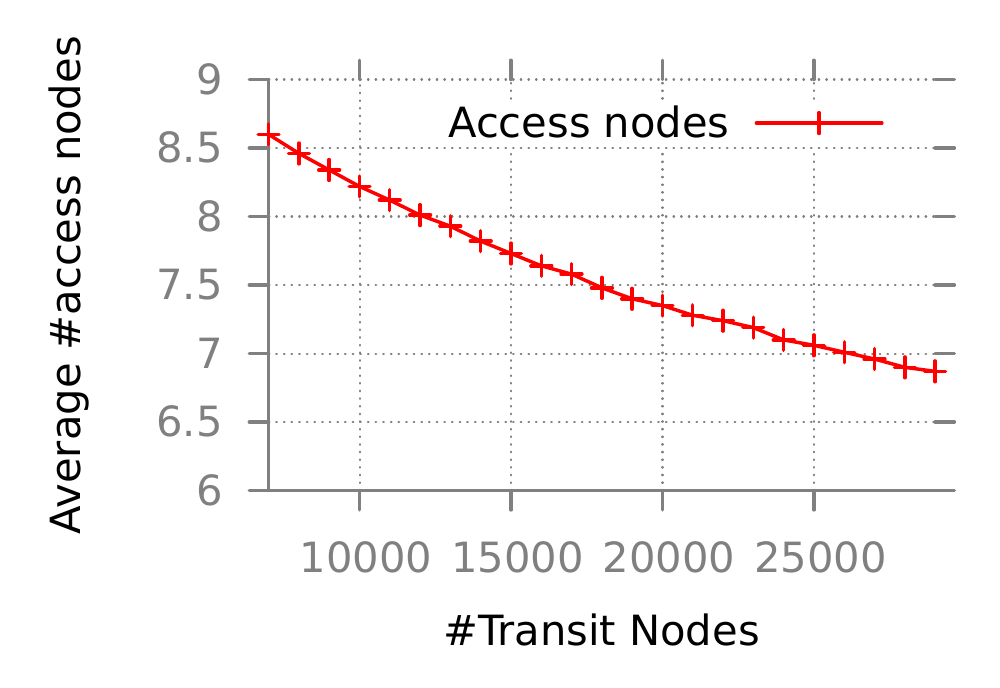}
\label{fig:plotAN}}
\caption{The average search space sizes (left) and access nodes (right) per graph node.}
\label{fig:nodesPerNode}
\end{figure}

The effect of transit node set size on space requirements is examined next.
Besides the underlying contraction hierarchy, the distance table, and access nodes, as well as the locality filter contribute to the space consumption.

The average numbers of nodes in the search space, Voronoi representatives and access nodes per node are plotted against varying sizes of $\mathcal{T}$.
For both values, the average between the respective forward and backward sizes is given, since the values are virtually identical.
We observe that these numbers fall as expected the larger the transit node set gets.
The results are plotted in Figure~\ref{fig:nodesPerNode}.
Obviously, the raw size of the CH is independent of $\vert\mathcal{T}\vert$ while the distance table grows quadratically. 
Space for access nodes slowly \emph{decreases} with $\vert\mathcal{T}\vert$ since the average number of access nodes decreases with more smaller local search spaces.
The same applies for the locality filter -- it needs less space although it gets more effective at the same time.

\begin{figure}[hbt]
    \centering
    \includegraphics[width=0.8\textwidth]{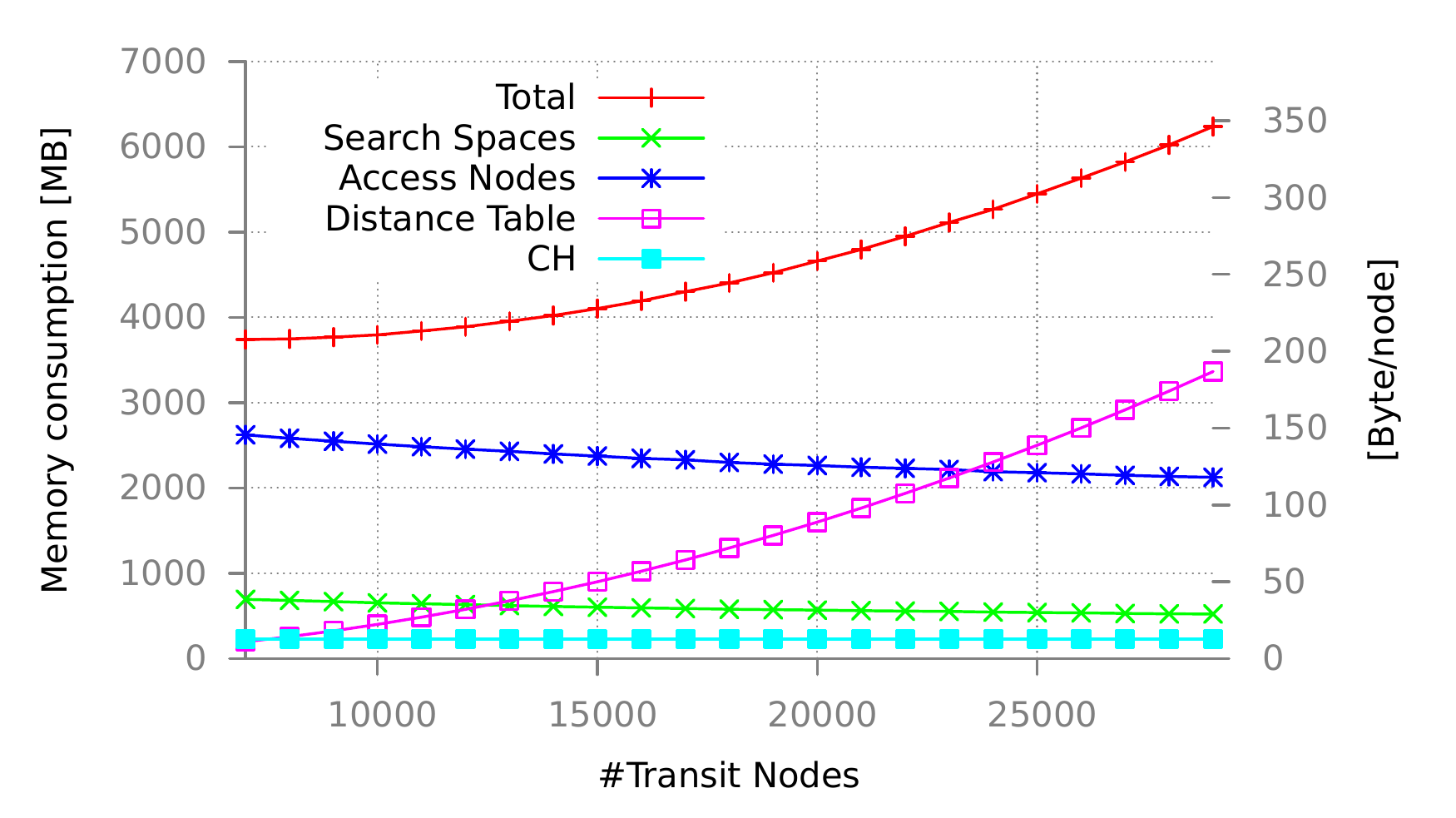}
    \caption{Memory consumption depending on $\vert\mathcal{T}\vert$ for an implementation without compression.}
    \label{fig:tnr-plotMemoryUncompressed}
\end{figure}

Figure \ref{fig:tnr-plotMemoryUncompressed} shows the relation between memory requirements and transit node set size.
Note that the implementation in this experiment does not merge search spaces or access node sets to give a clearer picture of the memory consumption of each part of our method.
We see that the main driver here is the size of the distance table which depends quadratically on the size of the transit node set.
Although, the average access node set decreases with an increasing transit node set size, it is not enough to compensate for the distance table.
We note that the space requirement of the search spaces is more or less constant over the entire parameter space.

\subsection{Results for Other Instances}\label{sec:tnr-otherInstances}

Further experiments are done with two additional instances.
The first one is the graph from above with distance metric (euro-dist).
The second test instance is an edge-expanded travel-time graph of Germany (ger-tc) extracted from OpenStreetMap\footnote{\url{http://osm.org}} at database timestamp \texttt{2012-12-10T19$\backslash$:23$\backslash$:02Z} extracted with the routines and car speed profile of Project OSRM\footnote{\url{http://project-osrm.org}}.
Edge-expansion implies that the graph explicitly models turn restrictions from the data.
Expanded graph nodes resembles undirected, i.e. unexpanded, edges from the input data, while expanded edges model allowed turns.
Note that U-Turns are explicitly forbidden.
The expanded graph of Germany is about twice as large as the unexpanded graph of Western Europe.
See Table~\ref{tab:tnr-graphs} for more details on the sizes of the instances.

\begin{table}[bth]
\caption{The graphs instance sizes alongside the number of edges in the respective CH search graph.}
\label{tab:tnr-graphs}
\centering
\begin{tabular}{lcrcrcr}
\toprule 
 instance & & nodes & & edges & & CH size \\
\midrule
 euro-time 	&  & 18\,015\,449 & & 22\,413\,128 & & 39\,256\,327 \\
 euro-dist 	&  & 18\,015\,449 & & 22\,413\,128 & & 44\,368\,351 \\
 ger-tc 		&  & 35\,024\,256 & & 43\,790\,686 & & 105\,617\,078  \\
\bottomrule 
\end{tabular}
\end{table}

\begin{table}[b]
\caption{The graphs instance sizes alongside the number of edges in the respective CH search graph.}
\label{tab:graphs}
\centering
\begin{tabular}{lcrcrcr}
\toprule 
 instance & & nodes & & edges & & CH size \\
\midrule
 euro-time 	&  & 18\,015\,449 & & 22\,413\,128 & & 39\,256\,327 \\
 euro-dist 	&  & 18\,015\,449 & & 22\,413\,128 & & 44\,368\,351 \\
 ger-tc 		&  & 35\,024\,256 & & 43\,790\,686 & & 105\,617\,078  \\
\bottomrule 
\end{tabular}
\end{table}

\begin{table}[h]
\caption{Experiments on different graphs. Values are scaled to match the same hardware.}
\label{tab:experimentsGraphs}
\centering
\begin{tabular}{lclcrrcrrr}
\toprule 
Graph 	& 		& $\vert\mathcal{T}\vert$ & & $\vert A\vert$ & $\vert\bar{S} \vert$ & Byte $/$ & time 		 &	& query \\ 
 	& 	& 	&  &  & & node & [\si{\minute}]  &	& [\si{\micro\second}] \\ 
\midrule
euro-dist	& & 10\,000 & & 18.0 & 22.1 & 440 & 32.4 & & 6.717 \\
	 		& & 15\,000 & & 17.1 & 18.7 & 424 & 28.9 & & 4.678 \\
	 		& & 25\,000 & & 14.5 & 14.7 & 456 & 26.1 & & 3.317 \\
\midrule
ger-tc		& & 20\,000 & & 9.11 & 14.66 & 278 & 18.9 & & 2.669 \\
			& & 40\,000 & & 7.78 & 11.85 & 383 & 16.9 & & 2.518 \\
			& & 50\,000 & & 7.37 & 11.01 & 476 & 16.6 & & 2.678 \\
\bottomrule 
\end{tabular}
\end{table}

Table~\ref{tab:experimentsGraphs} gives results for two further instances and three different sizes of the transit node set. Besides the European network with travel time metric used before (euro-time), we consider the same network with geographical distance metric (euro-dist), and a very detailed model of the German road network (ger-tc) based on OpenStreetMap data \footnote{Database timestamp \texttt{2012-12-11T19$\backslash$:00$\backslash$:02Z} } with explicitly modelled turns. 
This graph has 35\,024 nodes and 43\,790\,686 edges resulting in a CH with 105\,617\,078 edges.
U-Turns are explicitly forbidden and the turn restrictions from the input data were used.
The experiments on ger-tc were performed on the slightly slower Intel Xeon machine with more RAM. 
We determine scaling factors to compare the outcomes by running the algorithm with the European (time-metric) graph on that machine. 
The factors are 0.804 and 0.960 for preprocessing and query, respectively. 
The results of our experiments are scaled accordingly to the speed of the faster machine.

We see that the average number of access nodes as well as the average number of Voronoi representatives decrease with an increasing size of the transit node set.
The same holds true for preprocessing duration and query times.
The decrease in preprocessing duration is caused by likewise decreased search spaces in $V\backslash\mathcal{T}$.
Although, it takes longer to preprocess the distance table, the impacted is minor when compared to CH graph preprocessing.
Most nodes during a CH search are relaxed in the most upper portions of the hierarchy.
This implies that the search space exploration during preprocessing becomes faster, because the search spaces become smaller.

The same holds true for the query.
The larger the transit node set, the faster the queries become.
This is expected behavior, because of two reasons.
First, search spaces below the transit node set become smaller, as argued above.
Second, the number of local fallback queries decreases, because even more of the queries can be answered by table lookups.
Unfortunately, the distance table on the transit node set grows quadratically with the number of nodes.
Therefore, the overall space consumption increases again at some point, when the quadratic increase of the distance table can not be compensated by the falling average number of access nodes and Voronoi representatives.
We attribute this behavior that most edge relaxations during CH query happen in the highest portions of the hierarchy.
So, there is a point of diminishing returns, when the transit node set covers this dense portion of the hierarchy.
Note that the results shown in Table  \ref{tab:experimentsGraphs} are selected to reflect this observation.

If the input graph is not strongly connected, it may be that we end up with an empty set of forward or backward access nodes for some nodes.
In that case the minimum in Equation~\ref{eq:distance} minimizes over an empty set.
We define this minimum as $\infty$ correctly indicating non-reachability in case of a non-local query.
Similarly, non-existing paths between pairs of transit nodes will be detected during the precomputation and are indicating by a distance of $\infty$, too.
The search assigning Voronoi representatives may not reach all nodes and any unreached nodes are assigned to a dummy Voronoi region.

\begin{figure}[b!]
	\centering
	\includegraphics{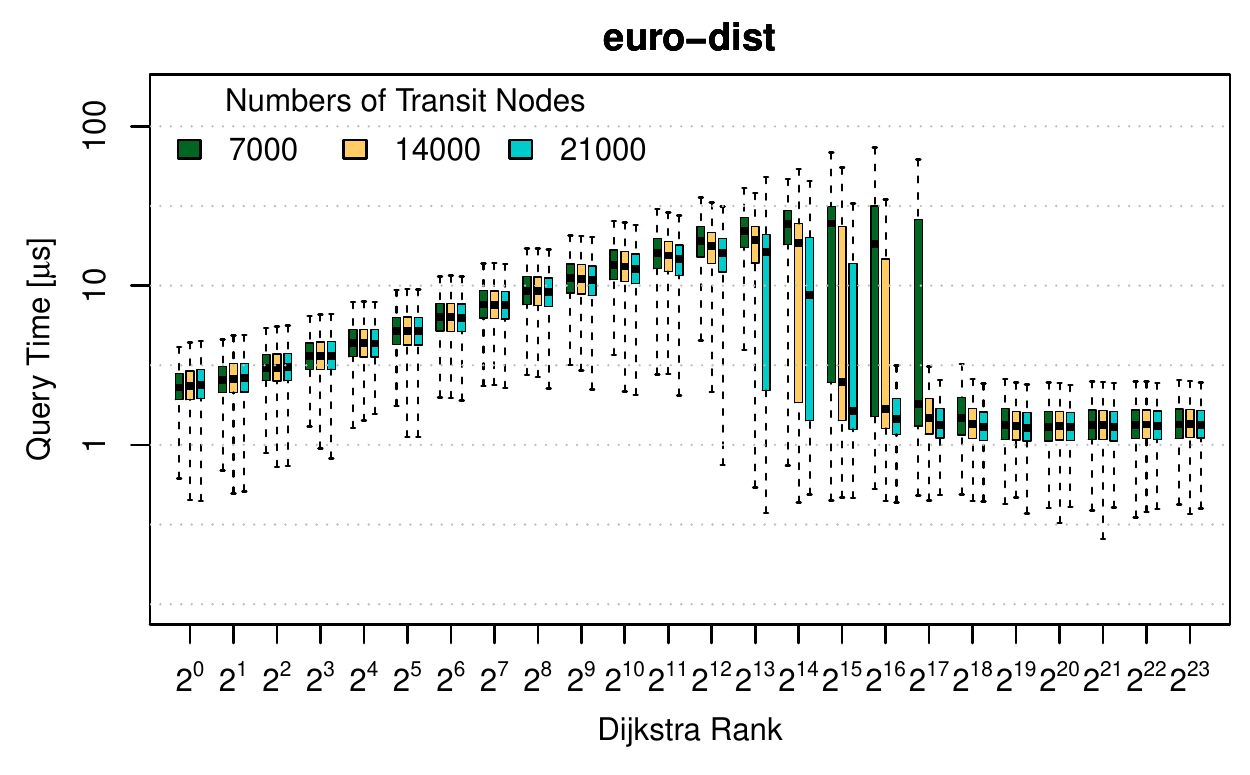}
    \caption{Rank plot for using $\vert\mathcal{T}\vert=\{7\,000, 14\,000,21\,000\}$ transit nodes.}
	\label{fig:rankPlots_dist}
\end{figure}

As to be expected from previous work,
we see that switching to distance metric is costly. 
The number of access nodes doubles  and accordingly space overhead also doubles. 
Since the number of table lookups is quadratic in the number of access nodes, the query time nearly quadruples.
On the positive side, the detailed model of the German graph, which is perhaps closest to state of the art routing applications, behaves similar to euro-time. 
The number of access nodes increases only slightly, and considering the larger graph size, the preprocessing time also remains moderate. 
This is an important difference to plain CHs where switching to a detailed graph model leads to significantly increased query time. 

\begin{figure}[t!]
\centering
	\includegraphics{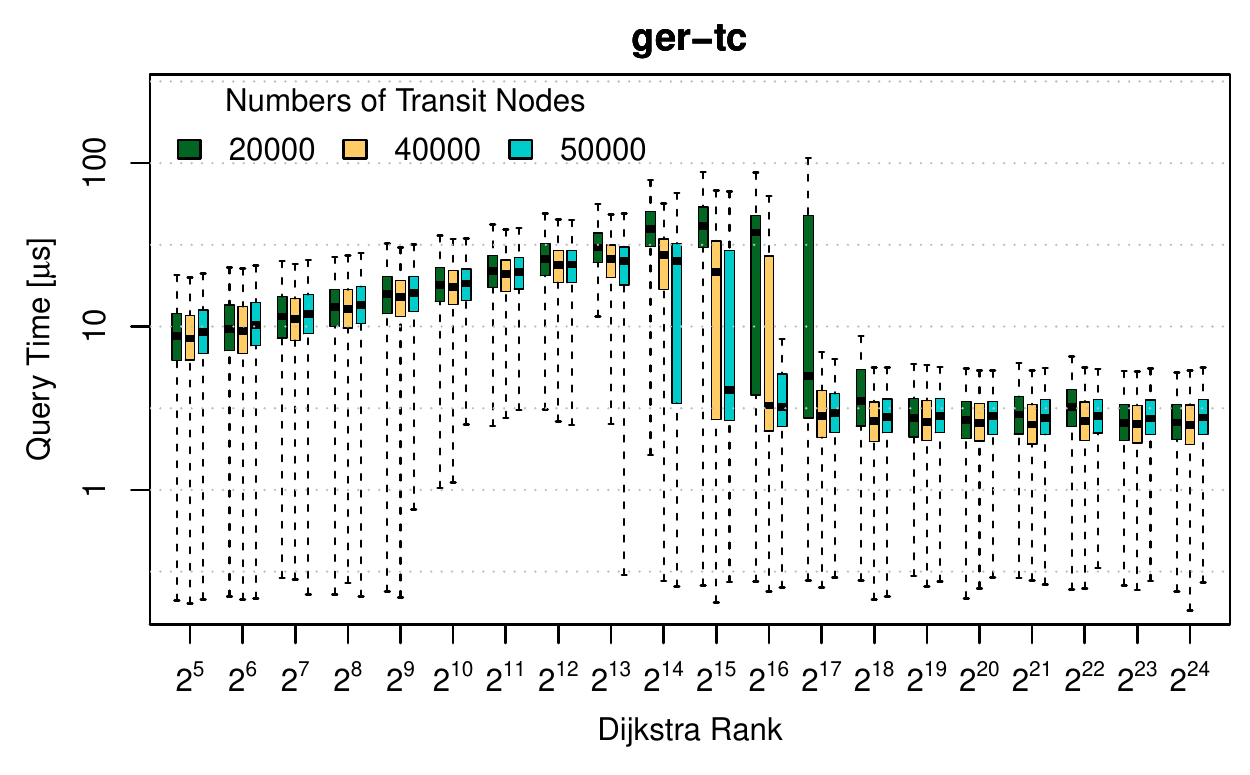}
	\caption{Rank plot for using $\vert\mathcal{T}\vert=\{20\,000, 40\,000, 50\,000\}$ transit nodes.}
\label{fig:rankPlots_ger}
\end{figure}

Figure~\ref{fig:rankPlots_dist} shows the rank plot for the \emph{euro-dist} instance for varying sizes of the transit node set.
Again, we observe that the parameter influences the threshold from which on shortest paths are covered by access nodes.
Figure~\ref{fig:rankPlots_ger} shows the result for the same experiment on the edge-expanded graph of Germany.
Note that we experimented on a much higher number of transit nodes on this instance, because it is much larger than the other instances.
Again, we observe a sharp cut-off from which on paths are covered by access nodes and that the value of $\vert\mathcal{T}\vert$ is a tuning parameter when this cut-off occurs.
Most interestingly, the queries seem to have a greater variance in the sense that there are outliers that have rather low query time.

\section{Further Improvements}

In addition to the previous experiments we identify a number of additional enhancements and use cases of our method.
Each of the following sub-section is work in progress at the time of writing.
Thus the results shall be treated as preliminary.

\subsection{Pruning with Arc Flags}\label{sec:tnr-af}
The experimental evaluation of Section \ref{sec:tnr-experiments} shows that the majority of the query time is spent in table lookups and only to a lesser extent in the locality filter.
In this Section, we analyse how further pruning using arc flags could be applied to achieve a sub-microsecond distance oracle on a fictive 2 GHz CPU.
Thus, while the following numbers are encouraging, they have to be taken with a grain of salt.
First, we briefly explain the layout of the query and reach for previous work by Bauer \etal \cite{bdsssw-chgds-08}, Delling \cite{d-earpa-09} and Abraham \etal \cite{dgnw-phast-12} to conduct the preprocessing.
Second, we note that the cost of a L1 cache miss is about 10 cycles and the cost of a L3 cache miss is about 100 cycles on a modern memory architecture.
A L1 cache hit is accounted for by a single nano-second.
On a 2 GHz machine this amounts to 5 and 50 nanoseconds, respectively.
These numbers were determined experimentally by Luxen and Schieferdecker \cite{ls-dmlca-12} when researching the cost associated with low-level memory accesses.

The TNR query can be sped up by using arc flags as previously reported, e.g. \cite{bdsssw-chgds-08}.
It is very similar to traditional arc flags.
Instead of partitioning the entire input graph only the core induced by the transit nodes is preprocessed.
Before running the table lookups for each and every pairwise combination of access nodes, a small set of auxiliary data in cache is queried, if the table lookup and the associated expensive cache miss is necessary at all.

If we partition the overlay network induced by the transit nodes into 48 regions, like \cite{d-earpa-09}, then this would require 96 bits (uncompressed) for each node to store forward and backward flags.
This totals in less than 120 KBytes of additional information for the exemplary 10\,000 transit nodes from Section \ref{sec:tnr-experiments}, which easily fits into the L2 cache of any modern processor.
Further, we assume that the access nodes for each node have been sorted previously.
The entire pruning table can not be scanned sequentially, but only a cache line of 32 bytes at a time.
But since the data is sorted it is fair to assume that six fetch operations into L2 cache suffice.
This accounts to our fictive runtime with about 5 nanoseconds for each L1 cache miss and a single nanosecond for each \emph{AND}-operation.

The following numbers are exemplary for the \texttt{euro-time} graph.
If we have $6.1$ Voronoi access nodes on average, we expect to check $6.1\times6.1$ pruning flags in total.
As Bauer \etal report, the remaining number of table lookups for TNR-AF is $3.1$ which will cost us a L3 cache miss in the worst case.
Thus, we account for an additional 55 nanoseconds for each such access.
If a local query CHASE query costs $6.1$ microseconds on average \cite{dgnw-phast-12} and is conducted for $0.581\%$ of all queries, then the local searches account for roughly 100 nanoseconds on average.
This amounts to an expected total query time on a fictive 2 GHz CPU of
$$250 \text{ns} + 6\cdot 5 \text{ns} + (6.1)^2\cdot 1 \text{ns} + 3.1\cdot 55 \text{ns} + 100 \text{ns} \simeq 590 \text{ns} , $$
assuming that a single Voronoi-Locality filter invocation costs about 250 nanoseconds.
This is a conservative estimate since it does not account for any SIMD tuning opportunity.

For CHASE preprocessing, recent work of Abraham \etal \cite{dgnw-phast-12} gives an efficient algorithm, called \emph{PHAST}, to compute arc flags on our test instance in mere 14 minutes of preprocessing including CH construction.
The additional memory overhead amounts to roughly 600 MB.
Further, we make the (simplifying) assumption that computing the arc flags for the transit node overlay graph can be computed without much additional cost.
Thus, we conclude that it is possible to construct a sub-microsecond distance oracle in about 15 minutes.

\subsection{Multilayer CH-based TNR}

In \cite{bfmss-itcsp-07,s-rprn-08} local queries were also handled fast by introducing additional \emph{layers} of secondary and tertiary transit nodes. Because of the higher quality of our transit node sets and because CH-routing is faster than highway-hierarchy routing, this becomes less important than in \cite{bfmss-itcsp-07}, yet, at least a secondary layer would be useful to further reduce the query times observed in Section \ref{sec:tnr-experiments}.

So, consider a set $\mathcal{T'}\supset\mathcal{T}$ of secondary transit nodes.\footnote{The construction can be generalized to more than two layers easily.} 
In our CH-based setting this will just be the highest $k'$ nodes from the CH for some $k'>|\mathcal{T}|$.
An $s$--$t$ query will first invoke the top-level locality filter $L$. If this filter comes out positive, the secondary locality filter $L'$ will be invoked.
$L'$ can be implemented as before using the CH search spaces from $s$ and $t$, this time staying below level $k'$. Note that this information is gathered as a side effect of gathering the search space information for $L$.
Similarly, we can obtain the secondary access nodes in a way analogous to the
top level access nodes by analyzing the CH search spaces below level $k'$.
When both $L$ and $L'$ are positive, we perform a CH query which 
is now even more local than in the single-layer case.

The main challenge is handling the case when $L$ is positive yet $L'$ is negative and we have to use a lookup table $D_{\mathcal{T}'}$ for routing in the secondary arterial network. Note that we cannot simply store a complete distance table for $\mathcal{T'}\times\mathcal{T'}$ since that would be too big and since then there would be no point in having the $\mathcal{T}\times\mathcal{T}$ distance table.\footnote{An outline of multilayer CH-based TNR in \cite{Bast11} misses this important point.} Rather, we have to precompute distances $\mu(u,w)$ where $u,w\in\mathcal{T}'$ and where no node in $\mathcal{T}$ lies on this path (the latter case can be covered by using $D_{\mathcal{T}}$). Note that only nearby nodes in $\mathcal{T}'$ will require such entries. The sparseness of $D_{\mathcal{T}'}$ can be accommodated by using a hash table rather than a two-dimensional array.  The entries of $D_{\mathcal{T}'}$ are computed using a refinement of the many-to-many technique from \cite{ksssw-cmmsp-07,gssv-erlrn-12}. The backward searches upward from nodes $w\in\mathcal{T'}$ store shortest path information to $w$ in the nodes below level $k$ they reach. The forward searches upward from nodes $u\in\mathcal{T'}$ use this information to generate candidate entries for $D_{\mathcal{T}'}$.  These entries are validated by checking whether they are actually shorter than the best path using the top level arterial network.
\subsection{TNR Based Many-to-One Computations}\label{s:manyToOne}

Consider a scenario where we have to find many $s$--$t$-shortest path distances for a fixed $t$. The case for fixed $s$ works analogously. For example, this might be interesting for generalizations of A* search to multiple criteria where we can use exact single-criteria searches for pruning the search space. Although we can use Dijkstra's algorithm here (one backward search from $t$) for precomputing all single-criteria distances, this is expensive when the A* search touches only a small fraction of the nodes. 

The idea is to precompute $v$--$t$-distances for all transit nodes $v\in\mathcal{T}$ and to store them in a separate array $T$. 
This can be done using $|A^\downarrow(v)|\cdot|\mathcal{T}|$ table lookups accessing only $|A^\downarrow(t)|$ rows of the distance table.
Note that $T$ is likely to fit into cache. 
For a fast locality filter specialized for a particular target node, one can employ highly localized backward search from t, explicitly precomputing the nodes requiring a local query.

Ideally, we would like to precompute $\mu(s, t)$ for all source nodes $s$ which require a local query. 
In principle, this can be done using a single backward Dijkstra search from $t$. 
Rather than exploring the full graph, this backward search can stop when its search space is covered by transit nodes. 
The locality filter then uses a single array $D$ initialized to $\infty$. 
The backwards search sets $D[v]$ to $\mu(v, t)$ when it settles a node $v$ such that the path from $v$ to $t$ is not covered by a transit node (the Dijkstra search has to propagate this coverage information).
This corresponds to the conservative approach described in \cite{s-rprn-08} for highway node routing. 
However, for instances with very long edges such as ferry connections this conservative approach can take a lot of time.
There a several alternatives. 
A simple one is to use the stall-on-demand technique for covering search from \cite{s-rprn-08}, which should not be confused with the related technique used for a CH query.
We can also use the stall-in-advance technique from \cite{s-rprn-08} which might be faster yet results in a one-sided error for the locality filter.

For a non-local query we compute the distance
$$\mu(s,t)=\min_{a\in A^\uparrow(s)}d_{A^\uparrow}(s,a)+T[a].$$
Note that this takes time linear rather than quadratic in the number of access nodes and only incurs cache faults for scanning $A^\uparrow(s)$.
Preliminary experiments indicate that this method can yield an order of magnitude in query time improvement compared to TNR (to around 100ns for the European instance).
\section{Conclusions and Future Work}\label{sec:tnr-conclusion}

We have shown that a very simple implementation of CH-TNR yields a speedup technique for route planning with an excellent trade-off between query time, preprocessing time, and space consumption.  In particular, at the price of twice the (quite fast) preprocessing time of contraction hierarchies, we get two orders of magnitude faster query time.  Our purely graph theoretical locality filter outperforms previously used geometric filters. To the best of our knowledge, this eliminates the last remnant of geometric techniques in competitive speedup techniques for route planning.  This filter is based on intersections of CH search spaces and thus exhibits an interesting relation to the hub labelling technique.

When comparing speedup techniques one can view this as a multi-objective optimization problem along the dimensions query time, preprocessing time, space consumption, and simplicity. Any Pareto-optimal (i.e. non-dominated) method is worthwhile considering and good methods should have a significant advantage with respect to at least one measure without undue disadvantages for the other dimensions. In this respect, CH-TNR fares very well. Only hub labelling achieves significantly better query times but at the price of much higher space consumption, in particular when comparable preprocessing times are desired. 
Moreover, simple variants of hub labelling have even worse space consumption and less clear advantages in query time. When looking for clearly simpler techniques than CH-TNR, plain CHs come into mind but at the price of two orders of magnitude larger query time and a surprisingly small gain in preprocessing time. 

CH-TNR also has significant potential for further performance improvements. Our variant of CH-TNR focusses on maximal simplicity except for the Voronoi filter which is needed for space efficiency. But there are many further improvements that will not drastically change the position of CH-TNR in the landscape of speedup techniques but that could yield noticeable improvements with respect to query time, preprocessing time, or space at the price of more complicated implementation. We now outline some of these possibilities:

\paragraph*{Query time:}\label{sec:tnr-query-time}
In Section~\ref{s:manyToOne} we have seen that for the special case of many-to-one queries can be accelerated by another order of magnitude being the fastest known technique for this use case. But also the general case can be further accelerated. 
As in \cite{bdsssw-chgds-10} we could expect about twice faster queries by combining CH-TNR with arc flags for an additional sense of goal direction. The additional preprocessing time could be much smaller than in \cite{bdsssw-chgds-10} by using new CH based methods for fast parallel one-to-all shortest paths \cite{dgnw-phast-11}. Local queries can be accelerated by introducing additional layers as in HH-TNR.
Alternatively, we could use hub labelling for local queries.
This is still much more space efficient than full hub labelling and very simple since we need to compute local (sub-transit-node) search spaces anyway. 
This variant of CH-TNR can be viewed as a generalization of hub labelling that saves space and preprocessing time at the price of larger query times.

\paragraph*{Preprocessing time:} Besides CH construction the most time consuming part or CH-TNR preprocessing is the exploration of the sub transit node CH search spaces for finding access nodes and partition representatives. This can probably be accelerated by a top-down computation as in \cite{adgw-hhlsp-12}.
Note that using post-search-stalling we still get optimal sets of access nodes.
Finding Voronoi regions might be parallelizable to some extent since it explores a very low diameter graph.

\paragraph*{Space:} 
There are a number of relatively simple low level tuning opportunities here. 
For example, we can more aggressively exploit overlaps between forward/backward access nodes and search space representatives.
These ``dual use'' nodes need to be stored only in the access nodes set together with a flag indicating that they are also a region representatives.

We could also encode backward distances to access nodes as differences to forward distances. As in HH-TNR we could also encode access nodes of most nodes as the union of the access nodes of of their neighbors.

The region representatives stored by our graph Voronoi filter are virtually identical to the access nodes so that we only need to store a flag indicating whether an access node is also a region representative plus the few region representatives that are not access nodes also. 
In our experiments this would reduce space consumption by another $\approx15$ bytes per node. 

We already exploit that most forward access nodes are also backward access nodes but we could additionally exploit that the corresponding distances are also similar. 
Thus, we only need to store the forward distance and the difference to the backward distance. 
Preliminary experiments indicate that in the vast majority of cases, this difference fits into 16 bits in our experiments. 
The few remaining cases could be encoded as an escape value indicating that the true distance is stored in a small hash table of exceptional values. 
This would give another space saving of around $15$ byte per node.

Access nodes of a node $v$ are a subset of the union of the access nodes of their neighbors. 
Which access nodes are taken from where can be indicated by one small bit map for each neighbor. 
Hence it suffices to store access nodes for a dominating set of the nodes.
With all these measures together, space consumption of CH-based TNR could be pushed well below 100 byte per node.

\bibliographystyle{splncs}
\bibliography{submission-references,references}
\clearpage

\end{document}